\documentclass[journal]{IEEEtran}

\usepackage{cite}
\usepackage{amsmath,amssymb,amsfonts}
\usepackage{algorithmic}
\usepackage{graphicx}
\usepackage{textcomp}
\usepackage{xcolor}
\usepackage[]{hyperref}
\usepackage{subfigure}
\usepackage{booktabs}
\usepackage{multirow}
\usepackage{tabularx}
\usepackage{float}
\usepackage{siunitx}
\usepackage{algorithm}
\newcommand{\tabincell}[2]{\begin{tabular}{@{}#1@{}}#2\end{tabular}} 

\graphicspath{{./figures/}}

\begin{document}
%
\title{R2F: A Remote Retraining Framework for AIoT Processors with Computing Errors}

%

\author{Dawen~Xu,
        Meng~He,
        Cheng~Liu,
        Ying~Wang,
        Long~Cheng,
        Huawei~Li,~\IEEEmembership{Senior Member,~IEEE,} \\
        Xiaowei~Li,~\IEEEmembership{Senior Member,~IEEE,} 
        and Kwang-Ting~Cheng,~\IEEEmembership{Fellow,~IEEE}

\thanks{This article was presented in part at The 30th IEEE International Conference on Application-specific Systems, Architectures and Processors, 2019.}
\thanks{Dawen Xu, and Meng He are with both Hefei University of Technology, Hefei 230009, China and SKLCA, Institute of Computing Technology (ICT), Chinese Academy of Sciences (CAS), Beijing 100180, China.}
\thanks{Cheng Liu and Ying Wang are with SKLCA, ICT, CAS, Beijing 100180, China. (e-mail:liucheng@ict.ac.cn)}
\thanks{Huawei Li is with both SKLCA, ICT, CAS, Beijing 100180, China and Peng Cheng Laboratory, Shenzhen, 518055, China.}
\thanks{Long Cheng is with the School of Control and Computer Engineering, North China Electric Power University, Beijing 102206, China.}
\thanks{Kwang-Ting Cheng is with Department of Computer Science and Engineering, The Hong Kong University of Science and Technology, 999077, Hong Kong.}
}

\markboth{IEEE Transactions on Very Large Scale Integration (VLSI) Systems, ~Vol.~xx, No.~xx, xxx~2021}%
{Shell \MakeLowercase{\textit{et al.}}: Bare Demo of IEEEtran.cls for IEEE Journals}

\maketitle

\begin{abstract}
AIoT processors fabricated with newer technology nodes suffer rising soft errors due to the shrinking transistor sizes and lower power supply. Soft errors on the AIoT processors particularly the deep learning accelerators (DLAs) with massive computing may cause substantial computing errors. These computing errors are difficult to be captured by the conventional training on general purposed processors like CPUs and GPUs in a server. Applying the offline trained neural network models to the edge accelerators with errors directly may lead to considerable prediction accuracy loss.

To address the problem, we propose a remote retraining framework (R2F) for remote AIoT processors with computing errors. It takes the remote AIoT processor with soft errors in the training loop such that the on-site computing errors can be learned with the application data on the server and the retrained models can be resilient to the soft errors. Meanwhile, we propose an optimized partial TMR strategy to enhance the retraining. According to our experiments, R2F enables elastic design trade-offs between the model accuracy and the performance penalty. The top-5 model accuracy can be improved by 1.93\%-13.73\% with 0\%-200\% performance penalty at high fault error rate. In addition, we notice that the retraining requires massive data transmission and even dominates the training time, and propose a sparse increment compression approach for the data transmission optimization, which reduces the retraining time by 38\%-88\% on average with negligible accuracy loss over a straightforward remote retraining. 
\end{abstract}

%

\section{Introduction} 
\label{sec:intro}
Neural networks that enable intelligent or smart things are gaining increasing popularity in IoT devices \cite{fertilizing2020Loh}. They are usually both computing- and memory-intensive, and thus pose a great challenge to the general purposed processors (GPPs) in IoT devices with limited power budgets but real-time processing requirements in many applications such as obstacle detection in mobile robots, autonomous drones and vehicles \cite{kodali2017applications} \cite{lee2019neuro} \cite{venkataramani2019deeptools}. In this circumstance, numerous neural network accelerators closely coupled with a GPP, namely AIoT processors, emerge in IoT devices and the number grows rapidly over the years \cite{fertilizing2020Loh}. To ensure both low-power and real-time processing of the various neural networks, many AIoT processors are fabricated with newer technology nodes. For instance, Google Edge AI platform Coral is fabricated with \SI{7}{\nm} technology, and Navida Jetson Xavier adopts \SI{12}{\nm} technology. The small feature sizes of the transistors and higher clock frequency in these AIoT processors are more likely to be affected by the extreme environments and radiation, and greatly increase the probability of soft errors accordingly \cite{dixit2011impact} \cite{lin2020dad}, which can induce the computing errors and cause wrong prediction when the neural networks are deployed. The wrong prediction in many safety-sensitive applications such as autonomous driving, unmanned aerial vehicle, robotics, and engine failure prediction and diagnosis may lead to catastrophic consequences and losses. Although many classical fault-tolerant design techniques such as triple modular redundancy (TMR) can be utilized to mitigate the influence of the soft errors, they typically induce considerable overhead in terms of performance and power consumption, which contradicts with the real-time processing and low-power requirements of the typical AIoT applications. Thereby, lightweight yet effective fault mitigation techniques that will not incur neither notable performance penalty nor power consumption remain highly demanded.

Fortunately, we notice that, unlike generic applications, neural networks inherently involve redundancy and are more resilient to the computing errors \cite{nunez2018energy}, many neural network model optimizations like quantization and pruning essentially take advantage of this feature to obtain notable performance and energy efficiency improvement with minor inference accuracy penalty \cite{anwar2015fixed} \cite{merolla2016deep}. Hereby, a straightforward yet effective approach to mitigate the soft errors in the AIoT processors is to exploit the redundancy in the the neural network models with the retraining such that the computing errors along with the data can be learned by the retrained models. The retrained models that usually have the same sizes with the original models can be executed without performance penalty.

Retraining the neural network models to tolerate soft errors with marginal performance penalty and energy consumption overhead is promising for the AIoT processors, but it is non-trivial to conduct the retraining with existing deep learning frameworks such as Caffe \cite{jia2014caffe}, Tensorflow \cite{abadi2016tensorflow} and PyTorch \cite{paszke2017automatic}. First of all, existing frameworks typically have the entire training performed on the general purposed processors especially GPUs, but the AIoT devices with limited computing power and energy budgets usually are incapable of supporting the power-hungry GPUs and the neural network training directly. As a result, the training must be conducted on a powerful server either in the edge or in the cloud. On the other hand, it is rather difficult for the GPUs in the server to capture the influence of the soft errors on the neural network inference conducted on the remote AIoT processors exactly. The offline trained models with noise injection show only marginal model accuracy improvement when deployed in a different faulty environment according to our experiments in Section \ref{sec:motivation}. 

To address the above challenges, we propose a Remote Retraining Framework (R2F) for the fault-tolerant neural network models targeting at the resilient deployment on AIoT processors with soft errors. It takes the AIoT processors in the conventional training loop and exposes the on-site computing errors to the training framework such that the obtained models can learn the application data with the computing errors at the same time. More specifically, it has the forward propagation (FP) affected by the soft errors conducted on the remote AIoT processor and the backward propagation (BP) on the server. The iterative training process has the FP and BP conducted interchangeably. At the same time, the intermediate outputs of the FP need to be sent to the server for the gradient calculation and model updating. The updated model in BP needs to be transmitted to the AIoT processor for the inference in next iteration. However, there is still a lack of supporting gateware that enables the frequent communication between the AIoT processor and the server for the collaborated training. To that end, we define a set of high-level communication APIs to characterize the basic data transmissions between the remote AIoT processor and the server, and implement with the remote procedure calls integrated in ThingsBoard, a typical IoT software stack. With these APIs, the remote AIoT processor can be fitted to PyTorch for the remote retraining.

In addition, we observe that the retraining with R2F usually involves many training iterations and each iteration may include multiple batch processing. As a result, a large amount of data transmission is required between the AIoT processor and the server. Moreover, the size of the intermediate outputs of the batched inference can be much larger than that of the weights and the inputs. For instance, suppose the batch size is 16, the intermediate outputs of MobileNet (8bit fixed point) and ResNet50 (8bit fixed point) are $11\times$ and $19\times$ over the input images, and $30\times$ and $7\times$ over the weights respectively. At the same time, there is usually limited communication bandwidth between the AIoT processors and the server due to the resource constrain on the edge. Thereby, the frequent and large amount of intermediate data transmission poses a great challenge to the retraining. In this work, we propose a sparse increment compression scheme to reduce the data transmission. The basic idea is to apply TMR to the neural network execution on the AIoT processor to approximate the golden intermediate outputs of the inference. Then, we take the approximated intermediate outputs as the base and calculate the increments to the base. As the computing errors are rather sparse, the increments can be compressed effectively. When the compressed increment and the input features are transmitted to the server, the server can recompute the intermediate outputs with the transmitted input features and approximate the actual intermediate outputs with computing errors by adding the increments. With the proposed compression method, the intermediate data transmission can be greatly reduced and the retraining time can be cut down accordingly. 

On top of the on-site retraining time optimization, we also optimize the R2F for more elastic design trade-offs between the retrained model accuracy and model execution time on the AIoT processors with soft errors. Basically, we notice that straightforward on-site retraining shows limited model accuracy improvement under relatively higher fault error rate while TMR can be used to reduce the influence of soft errors and improve the model accuracy significantly. However, the TMR overhead is usually overwhelming especially for the AIoT processors with limited power budgets. In this circumstance, we apply a heuristic algorithm to select the most fragile layers and have them protected via TMR which is also utilized in the remote retraining. The neural network with partial TMR protection is implemented in R2F such that the retrained model accuracy can be improved with minor performance penalty even under higher fault error rate.


The contribution of this work can be summarized as follows:
\begin{itemize}
    \item We proposed R2F, an efficient remote retraining framework, to enable the collaborated neural network model retraining on both remote AIoT processors and the servers. It takes the remote AIoT processors with computing errors in the training loop and has the influence of the soft errors learned with the application data such that the retrained models can be fault-tolerant.
    
    \item We define a series of client-server communication APIs on top of typical IoT software stacks to facilitate the R2F implementation on a conventional training framework like PyTorch and a representative IoT software stack. Moreover, we further optimize R2F from the perspective of the retraining time and the model accuracy. Specifically, we propose a sparse increment compression to greatly alleviate the large data transmission overhead in retraining, and provide an elastic design trade-off between the model accuracy and performance penalty with an optimized partial TMR strategy. 
    
    \item According to our experiments on a set of typical neural networks, R2F reduces the training time by 38\%-88\% with the proposed data transmission optimization when compared to the baseline method. It achieves an elastic design trade-off between the model accuracy and the performance penalty with the proposed partial TMR protection. The top-5 model accuracy can be improved by 1.93\%-13.73\% while the performance penalty ranges from 0\%-200\% under high fault error rate.
    
\end{itemize}

The structure of this paper is organized as follows. Section 2 briefly introduces the related works on fault tolerant design of neural network models and accelerators. Section 3 analyzes the influence of soft errors on neural network accelerators and motivates the necessity of the neural network model retraining. Section 4 details R2F for resilient neural network execution on AIoT processors with soft errors. Section 5 introduces the proposed optimizations for R2F. Section 6 includes comprehensive experiments and evaluates R2F from different angles including accuracy improvement and training time. Finally, we conclude this work in Section 7.

\section{Related Work} \label{sec:relatedwork}
\subsection{AIoT Processors}
For the sake of both the low-power and real-time processing, neural network accelerators are increasingly utilized for the Artificial intelligence (AI) processing \cite{liao2017energy} \cite{verhelst2017embedded} in Internet of Things (IoT). They are usually closely integrated with a general purposed processor, and the integrated processor that enables AI on IoT devices is known as AIoT processor. Although numerous efforts have been devoted to the AIoT processor design especially the neural network accelerator design \cite{sze2017efficient} \cite{andri2016yodann}, it remains rather challenging to ensure the contradictory design goals of low energy consumption, high performance and prediction accuracy\cite{reagen2016minerva}  \cite{nabavinejad2020overview}. In this case, the reliability of the neural network processing on AIoT processors under soft errors further complicates the design. Conventional fault-tolerant techniques such as TMR that typically will induce substantial power consumption and performance penalty can not be used directly. More lightweight fault-tolerant approaches are highly demanded for the neural network processing on AIoT processors.

\subsection{Fault-tolerant Neural Network Processing}
Plenty of prior works have investigated the fault-tolerant processing of neural networks from many different angles \cite{reagen2018ares} \cite{xu2020persistent} \cite{li2017understanding} \cite{hanif2018error} \cite{salami2018resilience} \cite{mittal2020survey} \cite{shafique2020robust}. They can be roughly divided into three categories based on the fault-tolerant targets. Some of them attempt to develop fault-tolerant neural network architectures, some of them seek to harden the underlying hardware infrastructures while some of them adopt hybrid approaches that take both the hardware accelerators and the neural network models into consideration at the same time. They will be illustrated in detail in the rest of this subsection.

\subsubsection{Fault-tolerant Neural Network Models} 
According to the evaluation in \cite{reagen2018ares} \cite{xu2020persistent}, minor computing errors in the neural network execution may not necessarily cause the wrong prediction. Basically, neural networks are usually more resilient compared to generic applications because of the computing redundancy and activation functions that can mitigate the computing variations in the neural network models. Many prior works exploit this feature of neural network models to further improve the resilience of the neural network processing. Some of them mainly rely on the training by introducing additional noise or in-situ faults \cite{qin2017robustness} \cite{kim2018energy}, or by adding regularization or penalty terms \cite{he2018axtrain} \cite{torres2017fault} \cite{leung2016regularizer}, or adding constraints to the weights \cite{reagen2016minerva} \cite{schorn2019efficient}. Model retraining-based approaches are also applied specifically for the emerging yet imperfect RRAM-based neural network accelerators \cite{chen2017accelerator} \cite{xia2018fault}. Unlike these aforementioned works that will not change the structure of the neural network models, FTT-NAS \cite{ning2020ftt} \cite{li2020ftt} provides an end-to-end fault-tolerant neural network search to redesign the neural network architecture. The obtained neural network model is more fault-tolerant, but the accuracy still drops given higher fault injection eventually and it can still be improved with retraining. Liu et al. proposed to enhance the algorithm level error-resilience capability of DNN classifiers through a collaborative logistic classifier design by leveraging both asymmetric binary classification and an optimized variable-length “decode-free" scheme \cite{Liu2019fault}. Hoang et al. \cite{Hoang2020DATE} proposed to systematically define the clipping values of the activation functions that result in increased resilience of the networks against faults. The authors in \cite{gambardella2019efficient} analyzed the vulnerability of the different neural network layers, replicated the most fragile layers and scheduled the processing to minimize the influence of hard errors. 

\subsubsection{Fault-tolerant Neural Network Accelerator Architectures} 
To mitigate faults in the neural network accelerator caused by soft errors, an intuitive approach is to harden the neural network accelerator with conventional fault-tolerant circuit design techniques such as TMR. For instance, the authors in \cite{xu2019safety} proposed a block-based modular redundancy strategy to mitigate the faulty computing array blocks of the neural network accelerator. The work in \cite{ozen2019sanity} \cite{zhao2020algorithm} employed the spatial and temporal checksum to protect full connection and convolution layers in deep neural network models. The checksum-based approach originated from the algorithm-based fault tolerance for matrix-matrix multiplication enables both efficient error detection and correction \cite{hari2021making}. The authors in \cite{zhang2020sorting} proposed a parallel stochastic computing(SC)-based NN accelerator purely using bitstream computation by fully exploiting the superior fault tolerance of SC mainly for ternary neural networks. Li et al. \cite{li2020soft} proposed an error detecting scheme to locate incorrect Processing Elements (PEs) of the neural network accelerator and gave an error masking method to achieve fault-tolerance. Mahdiani et al. \cite{mahdiani2012relaxed} proposed to relax the fault-tolerance of the VLSI implementation by employing TMR to only the computation of the most important bits such that the hardware overhead is reduced and the critical path latency is improved without any accuracy penalty. Xu et al. \cite{dawen2020hybrid} has a dot-production unit to recompute the operations that are mapped to the faulty PEs in the 2D computing array without affecting the original dataflow. However, these techniques usually result in non-trivial overhead in terms of timing, area, and power consumption, which may fail the stringent performance and power consumption requirements of the AIoT applications. In addition, the accelerators need to be redesigned heavily, which can also be a barrier for the off-the-shelf products like Google Edge TPU.

\subsubsection{Hybrid Fault-tolerant Techniques} 
There are also a few works proposed to co-optimize the neural network models and the underlying neural network accelerators at the same time for higher resilience. The work in \cite{zhang2019fault} \cite{zhang2018analyzing} proposed to add additional bypass logic to PEs in the neural network accelerator and the output will be set to be constant such as zero for faulty PEs. On top of the accelerator with constant bypass, it further retrains the models to achieve higher prediction accuracy. The authors in \cite{abdullah2020salvagednn} proposed to add different bypass data paths to the PEs in neural network accelerators such that faulty PEs can be skipped and had the weights mapped to the faulty PEs pruned at the same time. Instead of directly pruning the weights, they reorganized the models to minimize the sum of the the saliency of the pruned neural network weights, which greatly alleviates the accuracy degradation. The authors in \cite{song2020itt} proposed a software and hardware co-design methodology to effectively preserve the classification accuracy of CNN with few on-device training iterations on RRAM-crossbars. Kim et al. \cite{kim2018matic} proposed an algorithm and hardware co-designed fault-tolerance framework called MATIC, which combines the characteristics of destructive SRAM reads with the error resilience of neural networks in a memory-adaptive training process. Ma et al. \cite{ma2019process} leverage the fault-tolerance of the neural network models to mitigate the faults caused by the process variation of the neural network accelerator with hardware bypassing and a novel weight transfer technique. In this case, the computing array of the neural network accelerator can run at higher frequency with limited accuracy drop.

In summary, retraining is usually applied to obtain fault-tolerant neural network models particularly for model-based and co-designed fault-tolerant approaches. However, prior works do not have the remote retraining overhead evaluated and ignore the overhead in IoT system with limited communication bandwidth.

\begin{figure}[b]
\centering
    \includegraphics[width=0.47\textwidth]{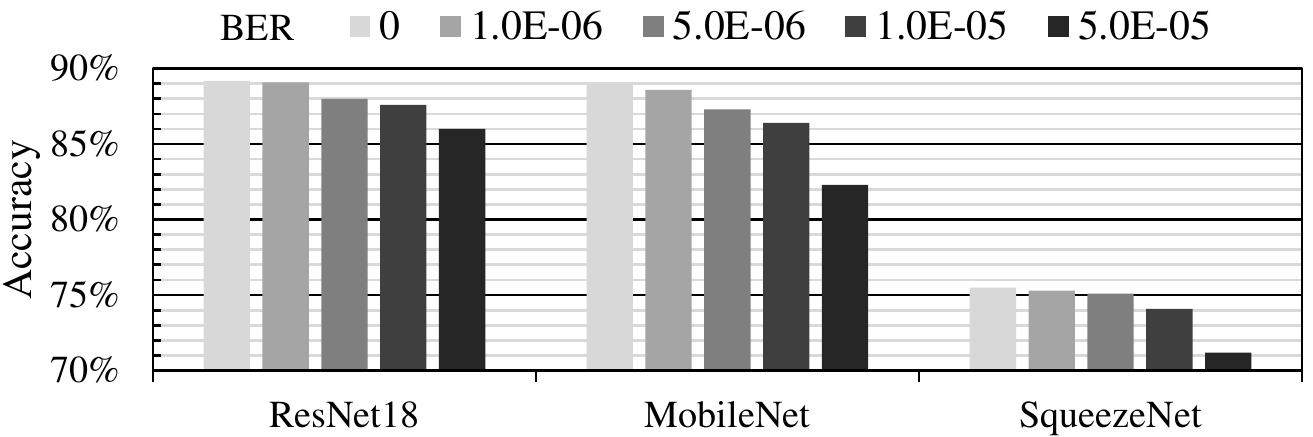}
    \vspace{-0.5em}
    \caption{Influence of the soft errors on the top-5 accuracy of different neural network models and bit error rate.}
\label{fig:influence}
\vspace{-1em}
\end{figure}

\begin{figure*}[tb]
\centering
    \includegraphics[width=1\textwidth]{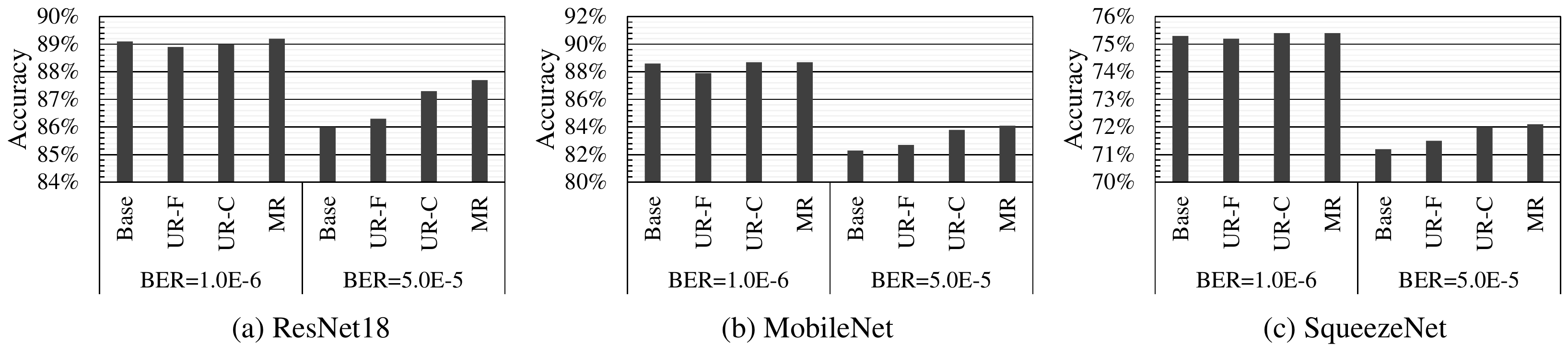}
    \vspace{-1.5em}
    \caption{The influence of the neural network retraining with matched and unmatched computing errors. Note that the BER used for training under 'UR-F' and 'UR-C' are $5.0\times10^{-5}$ and $5.0\times 10^{-6}$ respectively when the actual inference is conducted under $1.0\times10^{-6}$. BER used for training under 'UR-F' and 'UR-C' are $1.0\times10^{-6}$ and $4.5\times10^{-5}$ when the actual inference is conducted under $5.0\times10^{-5}$.}
\label{fig:noise}
\vspace{-1em}
\end{figure*}
\section{motivation} \label{sec:motivation}
In this section, we mainly investigate the influence of soft errors on the neural network prediction accuracy and effectiveness of retraining with computing errors, which motivates the proposed remote retraining framework.

\subsection{Influence of Soft Errors on Model Accuracy}
In order to evaluate the influence of soft errors on the neural network model accuracy, we have random bit errors injected to weights, inputs, outputs as well as the hidden states of the neural network models similar to the approach utilized in \cite{reagen2018ares}. We take three widely utilized neural network models, including ResNet18, MobileNet, and SqueezeNet pre-trained on ImageNet as the benchmark. All the models are 8bit fixed point. Note that the bit error rate (BER) represents the total number of bit errors over the total bit number of the data i.e. weights, inputs, outputs and hidden states of the neural network models. The experiment result is shown in \autoref{fig:influence}. It can be observed that the prediction accuracy of the neural network models drops little when BER is lower than $5 \times 10^{-6}$ though there are computing errors caused by the soft errors, which also demonstrates the intrinsic fault tolerance of the neural network models. Nevertheless, the accuracy drops rapidly when the BER reaches $5\times 10^{-5}$. Particularly, SqueezeNet drops by 6.04\% which is unacceptable for most of the accuracy-sensitive neural network models. Even for ResNet18 with the least accuracy drop, it also shows 3.72\% accuracy penalty which is non-trivial.

\subsection{Effectiveness of the Model Retraining on Soft Errors}
While it is promising to take advantage of the redundancy in neural network models with retraining to tolerate computing errors, we evaluate the effectiveness of the retraining on mitigating soft errors on a neural network accelerator in a remote AIoT processor. As the actual computing variation caused by the soft errors is not immediately available to the server and it is also difficult to model the exact variation on the server, we have the models retrained on the server with noise which is simulated via by injecting soft errors to the neural network accelerator with a different BER from that on the remote AIoT processor. Basically, we retrain the models with unmatched computing errors. Meanwhile, we also compare it with an on-site retraining which has the exact computing errors on remote AIoT processor transmitted to the server. The comparison is presented in \autoref{fig:noise}. Note that 'Base' refers to the model without retraining, 'UR' refers to unmatched retraining and 'MR' refers to the matched on-site retraining. Particularly for the unmatched retraining, we set two different bit error rate. One of them is far from the actual BER ('UR-F') and the other one is close to the actual BER ('UR-C'). It can be observed that retraining with unmatched BER that is far from the on-site situation i.e. 'UM-F' poses marginal model accuracy small improvements. In contrast, the retraining with matched retraining and unmatched retraining that is close to the on-site situation i.e. 'UR-C' exhibits much more significant accuracy improvement in general. 

In summary, retraining is generally beneficial to the prediction accuracy when the model is executed on a neural network accelerator affected by soft errors. Nevertheless, the retraining must have the computing errors caused by the soft errors considered and minor mismatch is acceptable. At the same time, the benefits will be dramatically undermined if the inference condition differs too much from the training condition.

\section{Remote Retraining Framework (R2F)} \label{sec:overview}
\subsection{Overview}
\begin{figure}
\centering
    \includegraphics[width=0.485\textwidth]{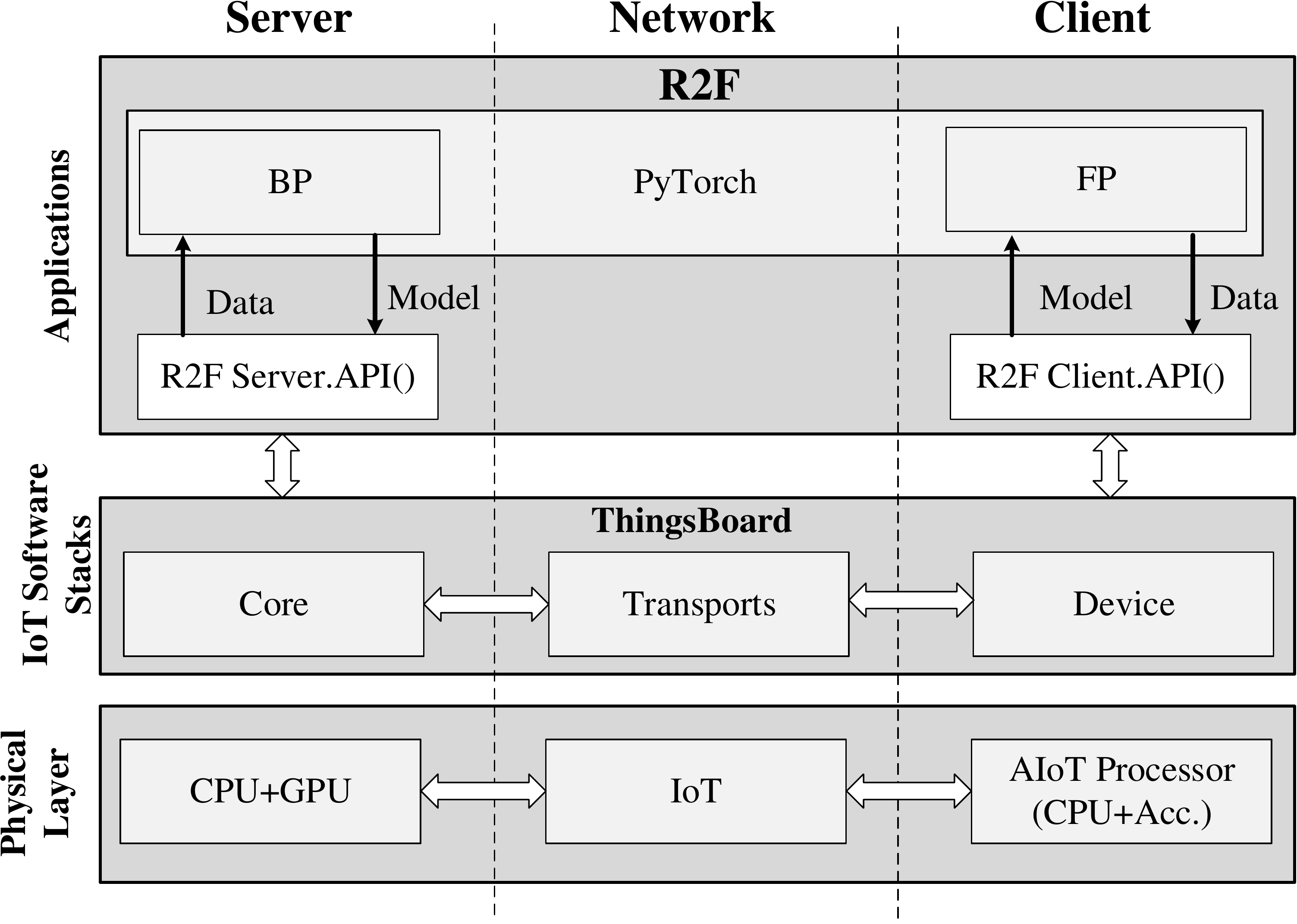}
    \vspace{-0.5em}
    \caption{An overview of the proposed remote retraining (R2F) framework on an AIoT system. It is essentially built on PyTorch and IoT software stacks, i.e. ThingsBoard, and integrates them with a series of client-server APIs.}
\label{fig:retrainoverview}
\end{figure}

To retrain a fault-tolerant neural network model for resilient execution on an AIoT processor, we opt to integrate the AIoT processor in the training loop of a conventional deep learning framework such that the computing errors caused by the soft errors can be learned and tolerated by the resulting neural network models, and develop a remote retraining framework (R2F) on top of PyTorch as shown in Figure \ref{fig:retrainoverview}. Unlike the conventional offline neural network training frameworks, it adopts a client-server computing diagram for the collaborated retraining between a remote AIoT processor and a server. On the server side, it reuses the conventional backward-propagation (BP) in PyTorch, but it needs to acquire the intermediate outputs of the neural network forward-propagation (FP) via a series of APIs denoted as \textit{R2F.Server.API()}. At the same time, it also needs to send the neural network models updated after BP to the client with the communication APIs provided by \textit{R2F.Server.API()}. On the client side, it receives the neural network models sent from the server and conducts the FP on the AIoT processor. Meanwhile, it has the intermediate outputs of the FP extracted and sent to the server with the \textit{R2F.client.API()}. 

Under the R2F, it is a typical IoT software stack and we adopt an open-source IoT framework called ThingsBoard (TB) in this work. TB enables cloud and IoT device connectivity via industry standard IoT protocols, such as, MQTT, CoAP, and HTTP and provides IoT oriented communication such as transport component. We take advantage of these communication facilities and wrap up them for the neural network training oriented communication i.e. \textit{R2F.server.API()} and \textit{R2F.client.API()}, which can be seamlessly integrated by PyTorch. The bottom layer is mainly the different hardware platforms. The server is typically equipped with both a powerful GPP and GPUs while the AIoT processor is usually configured with an neural network accelerator and a low-power GPP. They are connected with an IoT network supporting various communication protocols, such as LoRa, 802.15.4g, HSPA, and LTE Cat.4. When they are used in R2F, the AIoT processor collects input data from the sensors such as camera, and conducts the neural network models in a normal inference process while the server updates the neural network models based on the on-site inference outputs. 

\subsection{Intermediate Output Extraction}
Unlike the normal inference in which the outputs of the last layer of the neural network models are sufficient to obtain the prediction, the inference performed on the AIoT processor during the collaborated retraining needs to send the outputs of each neural network layer to the server for the gradient calculation and weight update in BP. However, many neural network accelerators are mainly optimized for inference without offering intermediate outputs. Outputs of some intermediate layers are completely stored in the on-chip buffer and directly consumed by the following neural network layer to reduce the accesses to the external memory. In this case, the accelerator need to add an optional data path to enable the intermediate output data write to the external memory on request. The intermediate output write can be done in parallel with the pipelined neural network execution. For some of the off-the-shelf neural network accelerators that do not support the intermediate output extraction, a more general approach to obtain the intermediate outputs is to divide the neural network models into sub models each of which includes a single neural network layer. When the sub models are compiled and executed sequentially, intermediate outputs can also be obtained, though it may take longer execution time. 

\begin{table}[tb]
\centering
\caption{Communication interfaces between a server and a client} \label{tab:api}
\begin{tabular}{|c|l|l|}
\toprule
\multirow{2}{*}{1} & API Name    & \textit{server.setAIoTMode(uint deviceID, uchar mode)}  \\ \cline{2-3} 
                   & Description & \tabincell{l}{
                   mode='inference', it sets the AIoT processor to\\
                    conduct inference only.\\
                   mode='training', it sets the AIoT processor to\\
                   dump intermediate outputs for retraining.} \\ \midrule
                   
\multirow{2}{*}{2} & API Name    & \textit{server.deployModel(uint deviceID, Model* model)}  \\ \cline{2-3} 
                   & Description & \tabincell{l}{
                   It sends the neural network model to the AIoT\\
                   processor for deployment.} \\ \midrule
                   
\multirow{2}{*}{3} & API Name    & \textit{server.getData(uint deviceID, uchar mode)}  \\ \cline{2-3} 
                   & Description & \tabincell{l}{
                   mode='inference', it receives outputs of the\\
                    inference sent from the AIoT processor.\\
                   mode='training', it receives both the intermediate\\
                   outputs and network outputs sent from the AIoT\\
                   processor.} \\ \midrule

\multirow{2}{*}{4} & API Name    & \textit{server.convertFloat2Int(float* fData, uchar* iData)}  \\ \cline{2-3} 
                   & Description & \tabincell{l}{
                   It converts the floating point model generated in BP\\
                   to fixed point for deployment.} \\ \midrule

\multirow{2}{*}{5} & API Name    & \textit{server.convertInt2Float(uchar* iData, float* fData)}  \\ \cline{2-3} 
                   & Description & \tabincell{l}{
                   It converts the received fixed point model to float\\
                   model for BP.} \\ \midrule

\multirow{2}{*}{6} & API Name    & \textit{client.sendAck(uint* deviceID)}  \\ \cline{2-3} 
                   & Description & \tabincell{l}{
                   It sends an acknowledgement to the server to ensure\\
                   the finish of the setup or commands.} \\ \midrule

\multirow{2}{*}{7} & API Name    & \textit{server.sendData(uint* deviceID, uchar mode)}  \\ \cline{2-3} 
                   & Description & \tabincell{l}{
                   mode='inference', it sends the neural network\\
                   outputs to the server.\\
                   mode='training', it sends both the intermediate\\
                   outputs and network outputs to the server on\\
                   request.} \\                   
\bottomrule
\end{tabular}
\end{table}

\subsection{Client-Server Communication Interface} \label{sec:communication-interface}
To enable the collaborated neural network model retraining with both an AIoT processor and a server that are far from each other, we define and implement a series of client-server APIs providing the high-level communication interfaces for the proposed retraining framework. With these APIs, we can adapt different types of AIoT processors and IoT software stacks to a unified retraining framework, i.e. R2F. These communication APIs are summarized in Table \ref{tab:api}. It can be classified as server APIs and client APIs running on the server and the clients respectively. The server APIs are mainly used to configure the mode of the remote AIoT processors, to deploy the updated neural network models, and to collect the inference results as well as the intermediate outputs sent from the remote clients. The client APIs are mainly used to response the different processing commands from the server and send the processing results to the server. There are two data conversion APIs used in the server side, because floating point is usually used in BP while 8-bit fixed point is mostly used in FP in the AIoT processor. More specifically, the received intermediate outputs will be converted to floating point for the float gradient calculation while the updated model will be quantized to 8-bit fixed point before it is sent to the AIoT processor for the deployment. Brief descriptions of all the APIs are listed in Table \ref{tab:api}.


\section{R2F Optimizations}
On top of the baseline R2F, we further optimize it from the following different angles. First, we optimize the communication time that dominates the on-site retraining time due to the limited uplink bandwidth of the AIoT processors. Second, we propose a partial TMR protection strategy to further improve the retrained model accuracy with minor performance penalty. Third, TMR is the basis of the aforementioned optimizations and its implementations are also optimized. They will be detailed in the rest of this section.

\subsection{Communication Optimization}
As discussed in prior section, R2F needs to transmit a large amount of intermediate outputs of the neural network models from the AIoT processors to the server, which requires both considerable time and power consumption due to the limited uplink bandwidth. Thereby, we seek to optimize the communication to enable efficient remote training. We notice that the computing errors caused by the soft errors can be effectively mitigated with classical TMR. Thus, we apply TMR to the neural network processing on the AIoT processor to obtain more accurate intermediate computing results of the neural network models. Since these results are close to the reference outputs, we take it as approximate reference outputs. Then, we calculate the increments of the intermediate outputs relative to the approximate reference outputs on the AIoT processor with soft errors. Since they are not significantly different, the incremental results includes a large number of zeros and can be compressed efficiently. Thereby, we can have the compressed increments instead of the intermediate outputs with computing errors sent to the server. At the same time, we also have the inputs of the neural network sent to the server. In the server, the reference outputs of the intermediate outputs can be re-calculated with the inputs. With both the increments of the intermediate outputs and the reference intermediate outputs, we can approximate the intermediate outputs of the neural networks in the AIoT processors affected by the soft errors. According to the motivation experiment in Section 3, these approximate outputs are still appropriate for the BP and on-site retraining on the server.

To support TMR on AIoT processors without hardware modification, we conduct temporal TMR directly. Basically, the neural network accelerator computes on the same inputs three times when it is set to be ‘training’ mode. The results of each output layer will be stored in memory. Then we have the general purposed processor to conduct the voter operations for each output. Since the data are sequentially stored, and the voting process can be improved with the vector processing unit inside the AIoT processor, the processing time of the voting stage is small compared to the inference time. 

\subsection{Critical Layer Protection}
Although the on-site model retraining improves the resilience of the neural network models to the soft errors, the prediction accuracy loss remains non-trivial under relatively higher fault injection rate. While the straightforward TMR on all neural network layers can greatly alleviate the influence of soft errors and extend the upper limit of the retraining method, it induces considerable performance penalty. Moreover, we notice that the computing errors on some layers of the neural network models may have distinct influence on the resulting prediction accuracy. Thereby, we select those layers that have the most significant influence on the neural network accuracy as critical layers and only have them protected with TMR to reduce both the model accuracy loss and the performance penalty. In addition, the implementation of the TMR is also consistent with that is mentioned in communication optimization.

In order to optimize the critical layer protection, we formulate the problem as follows. Suppose the target neural network includes $l$ layers, and $s$ layers are protected and the indices of the protected layer belong to a set $S$. The model accuracy of the neural network can be denoted as $A_S$. The computing overhead of each neural network layer is denoted as $O_i$ where $i$ represents the layer index. The design goal of the critical layer protection is to determine the set of the layers $S$ that need to be protected such that $A_S$ is maximized where the additional computing overhead relative to the original computing is less than $r$. To address the problem, we propose a heuristic algorithm to optimize the critical layer selection as illustrated in Algorithm \ref{alg:select}. The basic idea is to iteratively search the most critical neural network layer in a layer-wise manner. It continues the selection until the overhead of the partial TMR exceeds $r_{max}$.

\begin{algorithm}
	\caption{ Critical layer selection algorithm}
	\label{alg:select}
    \small
       \hspace*{\algorithmicindent} \textbf{Input:} A neural network with $l$ layers, the set of all the layer indices $L=(1,2, ..., l)$, the computing overhead of the $i$th layer is $O_i$ where $1 \leq i \leq l$. \\
       \hspace*{\algorithmicindent} \textbf{Output:} The total number of protected layers $s$ and the set of protected neural layer indices $S$ such that the $A_S$ is maximized and the normalized redundant computing overhead is no more than $r_{max}$. \\
	\begin{algorithmic}[1]
    \STATE $s = 0$, $S=\emptyset$
    \WHILE {$(r < r_{max})$} 
      \FOR{each $i\in (L\setminus S)$}  
        \STATE Measure the accuracy $A_{S\cup(i)}$
      \ENDFOR 
      \STATE Find $i$ that $A_{S\cup(i)}$ is maximized.
      \STATE $S$.append($i$), $s \leftarrow s+1$
      \STATE Calculate the normalized overhead $r=\frac{\sum_{i\in S}^{}O_i}{\sum_{i\in L}^{}O_i}$
      \IF{$r > r_{max}$}
	    \STATE $S$.remove($i$), $s \leftarrow s - 1$
	  \ENDIF
	  
    \ENDWHILE 
    \STATE Return $s$ and $S$

	\end{algorithmic}
 \end{algorithm}

With the proposed critical layer selection algorithm, the total number of design options that need to be evaluated in the search is $Sum(L, s)=\sum_{k=s+1}^{k=L}k=\frac{(L+s+1)\times (L-s)}{2}$. In contrast, a straightforward brute-force search requires to evalaute $C(L,s)=\frac{L!}{s!\times (L-s)!}$ design options. Take ResNet50 as an example, suppose $s=5$, each design option evaluation needs to conduct 1000 inference and takes around 16 seconds. The brute-force search requires to evaluate 2118760 design options and the total search time will be more than 392 days. Thus, it cannot be used in practice. The proposed search requires to evaluate 240 design options, which is 8820X faster and can be finished in around 64 minutes.

\subsection{TMR Implementation Optimization}
Although TMR is a classical redundancy approach, there are different methods to implement it on an AIoT processor. Since we will not change the architecture of the AIoT processors, a temporal TMR redundancy is used in this work. A straightforward TMR implementation is to conduct the neural network processing three times with the same input independently. Then the intermediate outputs from the three implementations are voted as the TMRed results. We name this approach as network-wise TMR (NW-TMR). While NW-TMR does not take the propagation of the computing errors across the different neural network layers into consideration, we propose to conduct the TMR in a layer-wise manner. Basically, the first layer of the neural network will be executed three times with the same input, and the outputs will be voted and the TMRed results will be used as the inputs of the next layer. This implementation is named as the layer-wise TMR (LW-TMR).  It is conducted iteratively until the end of the neural network model execution. We use the percentage of the identical outputs between the approximate intermediate outputs and the actual intermediate outputs as the output similarity evaluation metric. In order to avoid 
inefficient TMR that all the three data vary substantially, we only conduct the TMR-based compression to layers with higher similarity. The threshold of the output similarity can be changed for the different trade-offs between the training overhead and the retrained model accuracy. It will be evaluated in the experiment Section.

\section{Experiment} \label{sec:experiment}
\subsection{Experiment Setup}
\subsubsection{Hardware Platform} 
In the experiment, the server is configured with an Intel Xeon cpu E5-2699 v3@2.30GHz processor and 128 GB DRAM while we have a Raspberry Pi 3 Model B platform equipped with an ARM Cortex A53 processor and 1 GB memory as the AIoT processor. The hardware platform is mainly used to verify the functionality of the proposed R2F framework. Since the Raspberry Pi does not have a neural network accelerator integrated, we assume that a Eyerisis-like neural network accelerator simulated with Scale-Sim is used instead. The neural network accelerator is configured with $32 \times 32$ 2-D computing array, 1024 KB on-chip buffer. And the neural network models are executed with a classical weight-stationary dataflow. Moreover, the simulation-based neural network accelerator also facilitates the fault injection and analysis. The communication protocols used in IoTs greatly affect the bandwidth and even dominate the training time in R2F. While the different IoT protocols provide distinct bandwidth as shown in Table \ref{character} and they are usually utilized for different domains of IoT applications, 802.15.4g (802) and HSPA with moderate bandwidth and power consumption are more likely for the low power AIoT applications, and they are evaluated in the experiments.

\begin{table}
    \centering
  \caption{Typical IoT communication protocols}
  \vspace{-0.5em}
  \scriptsize
  \label{character}
  \begin{tabular}{cccc}
    \toprule  
      Technology & Typical Applications & \multicolumn{2}{c} {Bit Rate}  \\
       &  &Downlink&Uplink  \\
    \midrule
      LoRa & smart street lights and meter
                   & \multicolumn{2}{c} {50 kbit/s} \\
      802.15.4g & remote monitoring, industrial control
      & \multicolumn{2}{c}{800 kbit/s}  \\
      HSPA & shared payment, wearable device & 21.1 Mbit/s & 5.76 Mbit/s \\
      LTE Cat.4 & smart medicine, autonomous driving& 150 Mbit/s &50 Mbit/s\\
  \bottomrule
\end{tabular}
\vspace{-1em}
\end{table}

\subsubsection{Software}
we use ThingsBoard as a representative IoT framework and PyTorch as a typical deep learning framework for R2F. In order to compress the increments of the intermediate inference outputs on the ARM processors for efficient data transmission, we utilized the optimized LZ4 implementation in \cite{yan2020lz4} that offers fast lossless compression in the experiments.

\subsubsection{Fault Injection} 
Soft errors are randomly distributed to all the memory cells of the neural network accelerator including the register files and the on-chip buffers. When a memory cell is affected by a soft error, the bit in the memory cell will be flipped. The soft error rate i.e., bit error rate (BER) is defined as the ratio of the bit faults over the total number of the memory cells. As the soft errors in the register file and input/output/weight buffers essentially affect the input and output of each MAC (multiply-accumulate), we have the influence of the soft errors in the neural network accelerator converted to the random bit flip of both the input and output of each basic MAC in the neural network processing similar to prior works \cite{reagen2018ares} \cite{ozen2019sanity} \cite{ning2020ftt} \cite{li2020ftt}. Basically, bit errors are randomly injected to the input features, weights, hidden states, and output features during the neural network execution. It mainly evaluates the soft errors in algorithm level and does not take the soft errors in the controlling logic into consideration. This algorithm-level fault analysis strategy is also demonstrated in \cite{he2020fidelity}. In this case, BER denotes the number of bit errors relative to the total bit number of the weights, inputs, hidden states, and output features. In the experiments, we investigate a broad range of BER setups starting from $1.0\times 10^{-6}$ to $1.0 \times 10^{-4}$. In addition, we focus on the fault tolerance of neural network models which are usually deployed on a deep learning accelerator in an AIoT processor, and assume that the GPP processor is reliable.

\begin{table}
    \centering
  \caption{Neural Network Benchmark}
  \vspace{-0.5em}
  \scriptsize
  \label{Benchmark}
  \begin{tabular}{cccccc}
    \toprule  
      Network & ResNet18 & ResNet50 & MobileNet & ShuffleNet & SqueezeNet \\
    \midrule
      Model Size & 1.2 MB & 24.3 MB & 3.3 MB & 1.3 MB & 1.2 MB \\
      \midrule
      \# of Layers & 20 & 53 & 52 & 57 & 26 \\
  \bottomrule
\end{tabular}
\vspace{-1em}
\end{table}

\subsubsection{Neural Network Benchmark} 
In the experiment, we take five typical lightweight neural network models including ResNet18, ResNet50, MobileNet, ShuffleNet, and SqueezeNet utilized as the neural network benchmark. All the models are 8-bit fixed point models and pre-trained for ImageNet dataset. Details of these neural network models can be found in Table \ref{Benchmark}. The number of the convolution layers ranges from 20 to 57. The sizes of the neural network models ranges from 1.2 MB to 24.3 MB. In the experiments, we select 50000 images from ImageNet for the retraining, set the epoch to 1 and the batch size to 16.

\begin{figure*}[tb]
\centering
    \includegraphics[width=0.99\textwidth]{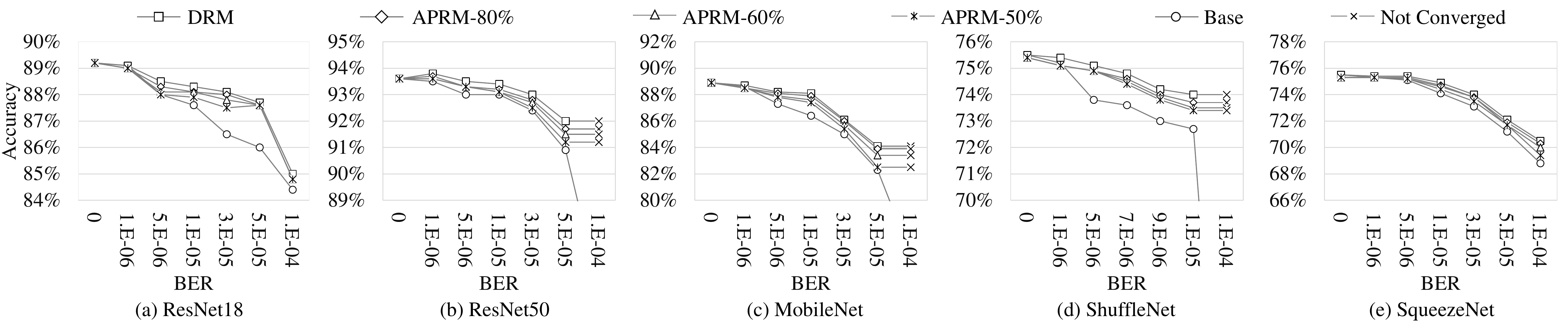}
    \vspace{-0.5em}
    \caption{The achieved prediction accuracy of the neural network models trained with data compression optimization under different output similarity thresholds.}
\label{fig:accuracy}
\vspace{0em}
\end{figure*}

\begin{figure*}[tb]
\centering
    \includegraphics[width=0.99\textwidth]{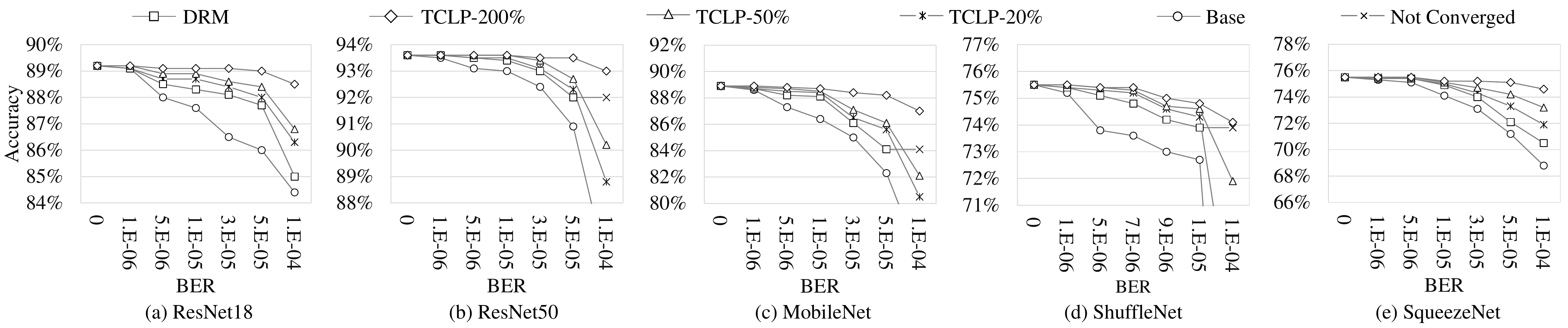}
    \vspace{-0.5em}
    \caption{The prediction accuracy of the retrained neural network models with different critical layer protection.}
\label{fig:protect}
\vspace{0em}
\end{figure*}

\subsection{Prediction Accuracy Improvement}
In this section, we mainly evaluate the prediction accuracy of the retraining and have the different retraining approaches compared. The neural network models executed directly on the accelerators with soft errors are considered as the baseline (Base). R2F puts the remote neural network accelerator into the training loop such that the retrained model can be fault-tolerant. The directly retrained model is noted with (DRM). While the direct retraining with R2F requires a large amount of intermediate data transmission particularly from the AIoT processor to the server, we further apply the TMR-based retraining, which has the compressed sparse increments rather than the intermediate outputs transmitted to the server for the retraining. The retrained model is denoted as an approximate retrained models (APRM). For the TMR-based retraining, we also explore the trade-offs between the percentage of the TMRed layers and the data transmission reduction. Basically, when more layers are transmitted with the original intermediate outputs, the retrained model will be more close to DRM, but more data transmission is required. In this work, we use the percentage of the identical outputs between the approximate intermediate outputs and the actual intermediate outputs as a simple output similarity evaluation metric and we only conduct the TMR-based compression to layers with higher similarity. Note that the metric is obtained with an offline analysis on a single random input. We have three different similarity thresholds, including 80\%, 60\%, and 50\%, applied and evaluated in the experiments. The obtained models are denoted as APRM-80\%, APRM-60\%, APRM-50\% respectively.

The resulting model accuracy of the different retraining is compared in Figure \ref{fig:accuracy}. It can be observed that the prediction accuracy of the neural network model degrades gracefully at the beginning but drops rapidly when the BER rises to certain points according to the 'Base' curve. Basically, the neural network models are fault-tolerant to the errors with in certain limit, but the models suffer dramatic accuracy degradation when the faults reach to the limit. In contrast to the 'Base', the retrained models with on-site computing errors generally exhibit clear prediction accuracy improvement. While the improvement is trivial when the BER is low, it gets significant when the BER is relatively higher. The top-5 prediction accuracy of DRM improves by 1.93\% on average compared to 'Base' at the highest BER under which the models can still be retrained. However, the on-site retraining shows less accuracy improvement and even fails to converge for some of the neural networks under high BER. Particularly, the retraining that does not converge is denoted as 'X' in the figures. We argue that this may also be caused by the limited fault-tolerance of the neural network models. Generally, the retraining works in a certain limit of the BER and fails when the BER exceeds the limit. When we compare DRM and the different APRM methods, we notice that the prediction accuracy of the retrained neural network models show little difference and it confirms that TMR can be applied to reduce the data transmission in a large range of scenarios. 

In order to further enlarge the benefits of the retraining, we propose to apply TMR to a small set of the most critical neural network layers to avoid the substantial performance penalty of a conventional TMR while retaining the model accuracy as much as possible. We call this approach as TMR-based critical layer protection (TCLP). As we can adjust the number of the critical layers and the performance overhead to compromise with the prediction accuracy improvement, we have a set of TCLP implementations with different performance overhead evaluated. We set the redundant computing relative to the original neural network computing as the overhead metric. The different TCLP implementations are denoted as TCLP-200\%, TCLP-50\%, and TCLP-20\% respectively. TCLP-200\% essentially refers to the standard TMR implementations. In addition, we also have the DRM and 'Base' compared. The comparison is shown in Figure \ref{fig:protect}. It can be observed that the retraining on top of the conventional TMR-based protection, i.e. TCLP-200\% shows 13.73\% and 4.97\% accuracy improvement on average compared to both the 'Base' and 'DMR' particularly when the BER reaches to $1 \times 10^{-4}$. When the TMR overhead is constrained, the resulting model accuracy also shows clear improvement compared to the DRM but imposes much less performance overhead than the full TMR. For instance, TCLP-200\% shows 4.97\% accuracy improvement on average over the DRM with $1 \times 10^{-4}$ fault injection, while TCLP-50\% shows 2.01\% accuracy improvement over the DRM. On the other hand, the performance of TCLP-200\% is 4X lower compared to TCLP-50\%. Similar design trade-offs between the model accuracy improvement and the performance overhead is observed on all the neural network models in the benchmark. Moreover, the experiments also reveal that some of the neural network layers are more critical than the others and prioritizing these layers for TMR protection helps to achieve significant accuracy improvement with least performance penalty. In addition, these design trade-offs are supported in R2F and can be applied for different application requirements.

\subsection{Training Time Reduction}
\begin{figure*}[tb]
\centering
    \includegraphics[width=0.98\textwidth]{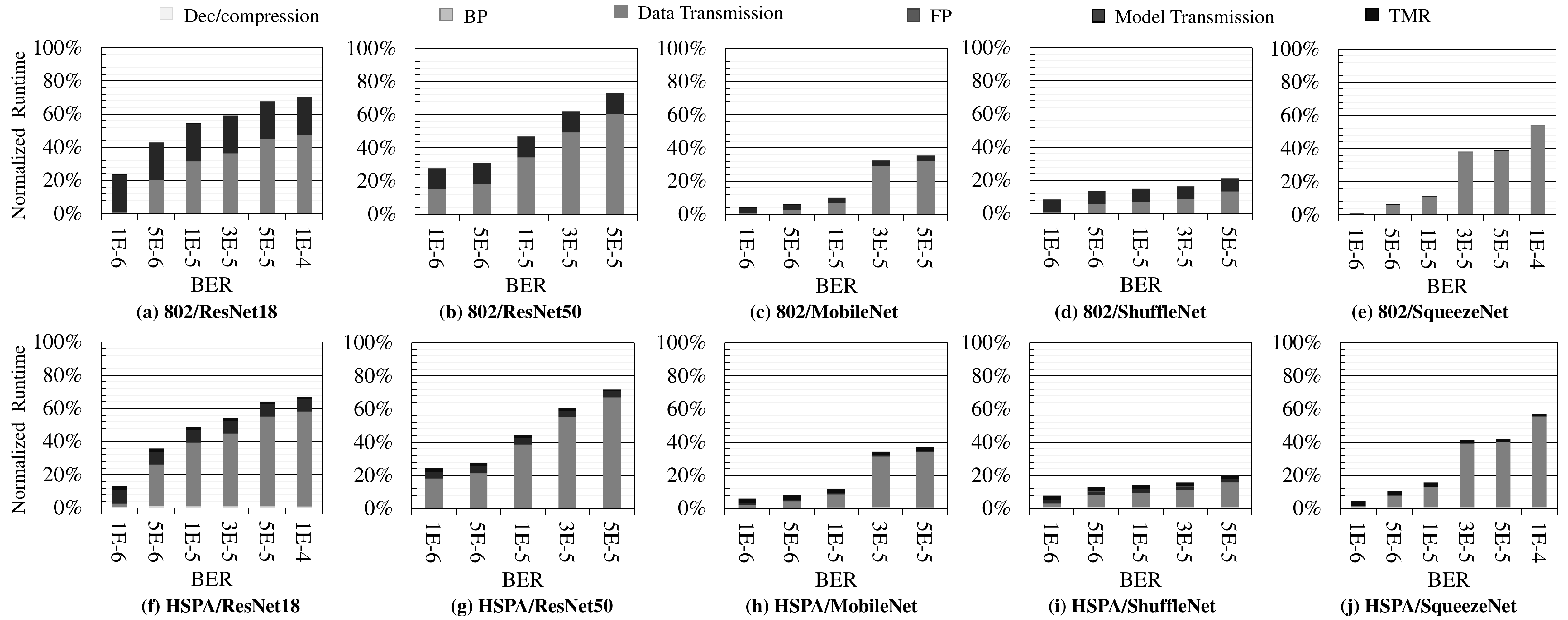}
    \vspace{-0.5em}
    \caption{The distribution of the retraining time normalized to the direct retraining. In particular, "HSPA/ResNet18" means that the AIoT processor communicates with HSPA protocol and the retrained neural network is ResNet18.}
\label{fig:runtime}
\vspace{-1em}
\end{figure*}

As discussed in prior section, the proposed TMR-based approximate retraining can greatly reduce the training time. We have the training time of the different neural network models under different IoT communication protocols decomposed and evaluated in Figure \ref{fig:runtime}. Note that the output similarity threshold in the experiment is set to be 60\% and will be discussed in the next subsection. The runtime of the retraining in R2F consists of six stages, including the forward propagation (FP), TMR processing of the intermediate outputs from FP (TMR), increment calculation, compression \& decompression (Dec/compression), data transmission from the AIoT processor to the server (Data Transmission), backward propagation (BP), and model transmission from the server to the AIoT processor (Model Transmission). To facilitate the comparison over the different neural network models and communication protocols, we have the training time normalized to that of DRM, which has the neural network intermediate outputs transmitted directly to the server. Note that 802 and HSPA listed in Table \ref{character} are used in this experiment. All the five representative neural network models listed in Table \ref{Benchmark} are evaluated. 

It can be observed that the data transmission dominates the retraining time when BER is relatively high. This is mainly because that the amount of the intermediate outputs of FP is usually large compared to the weights and it increases proportional to the batch sizes. Many of the network layers fail to meet the similarity threshold and require the direct data transmission at higher BER. On the other hand, the communication bandwidth provided by the typical IoT processors is limited and it further induces the large data transmission overhead. When the BER is lower, the majority of the neural network layers can benefit from the TMR-based data compression and the R2F training time is greatly reduced. In addition, we notice that the neural network models also have significant influence on the training time optimization. The normalized retraining time of ShuffleNet and MobileNet is much less compared to that of the rest neural network models. This may be caused by both the fault tolerance and sizes of the neural network models. Neural network models with smaller sizes and less computation are less probably to produce wrong computing results and thus are more likely to be optimized with the TMR-based compression in R2F. In contrast, ResNet18 and ResNet50 with much more computation may fail to meet the similarity threshold and more data needs to be transmitted directly. SqueezeNet is also as lighweight as ShuffleNet and MobileNet, but it is more sensitive to the soft errors as shown in Figure \ref{fig:accuracy}. As a result, it also requires considerable direct data transmission and consumes non-trivial time. In summary, the retraining time reduction is closely related with the BER. When the BER is low e.g. $1\times 10^{-6}$, the retraining time can be reduced by 88\% on average compared to that of DRM. When the BER is moderate e.g. $1 \times 10^{-5}$, the retraining time can be reduced by 73\% on average. When the BER is high e.g. $1 \times 10^{-4}$, the retraining time can be reduced by 38\% on average.


\begin{figure*}[tb]
\centering
    \includegraphics[width=0.99\textwidth]{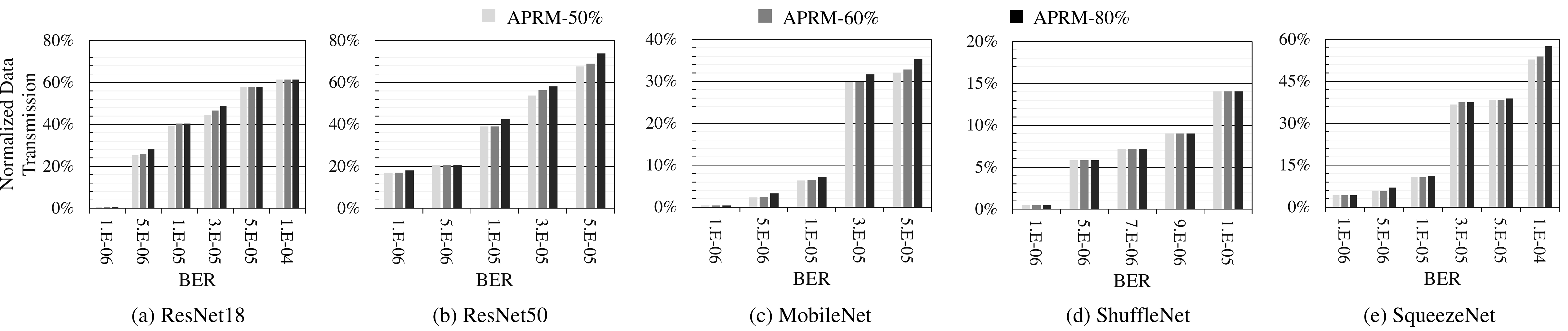}
    \vspace{-0.5em}
    \caption{The amount of data transmission required by R2F under different similarity metrics.}
\label{fig:data}
\vspace{0em}
\end{figure*}

\subsection{Design Option Tuning}
The data transmission reduction directly depends on the amount of the layers that can be approximated with TMR according to the output similarity metric. Thus, we have the amount of the data transmission required by the different approximate retraining approaches, including APRM-50\%, APRM-60\%, and APRM-80\%, normalized to that required by DRM and compared. The comparison is shown in Figure \ref{fig:data}. It can be observed that more data transmission can be reduced under lower BER when almost all the intermediate data can be effectively compressed with the TMR-based increment transmission. When BER rises, more neural network layers suffer considerable computing errors and these errors can not be mitigated with TMR. As a result, only a fraction of the neural network layers can be compressed under higher BER. Another observation from Figure \ref{fig:data} is that the output similarity threshold does not show dramatic data transmission variations. It is mainly because that the computing errors caused by the soft errors may aggregate and the number of the computing errors increases dramatically with the rising BER. As a result, the similarity metric is not proportional to the amount of the data transmission. The amount of the data transmission in the same neural network model retraining under the different BER also confirms this feature. In this work, we adopt 60\% as the similarity threshold to decide whether the outputs of a neural network layer will be transmitted directly or with the TMR-based increment compression.  

\begin{figure}
\centering
    \includegraphics[width=0.39\textwidth]{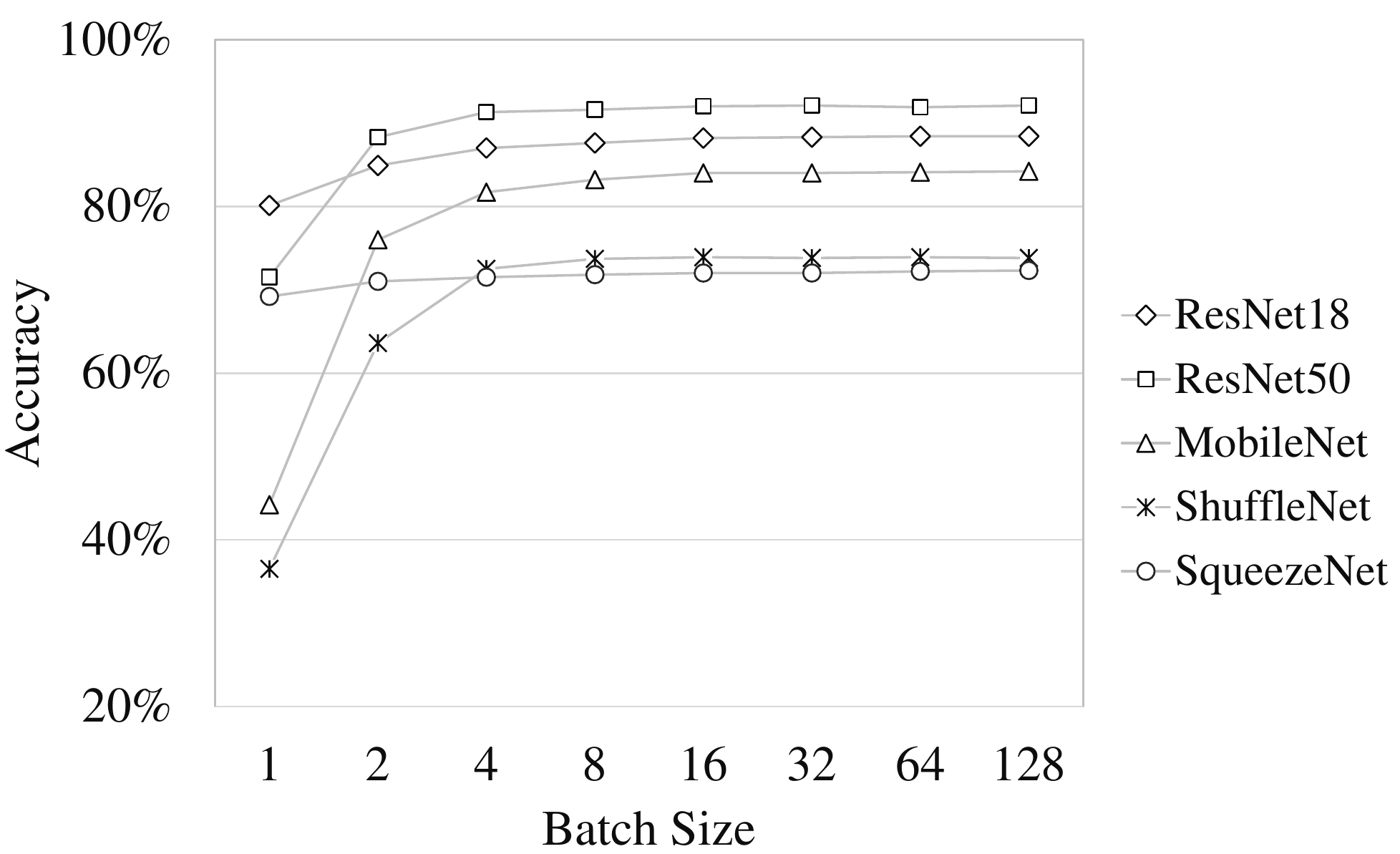}
    \vspace{-0.5em}
    \caption{The influence of batch size on the proposed R2F training.}
\label{fig:batchsize}
\vspace{-1em}
\end{figure}

The amount of the intermediate data transmission is proportional to the batch size and affects the R2F training time. Thus, we investigate the influence of the batch size on the retrained model accuracy. The experiment result is shown in Figure \ref{fig:batchsize}. It can be observed that larger batch size is generally beneficial to the model accuracy but the benefits roughly get saturated when the batch size reaches to 16. The main reason is that larger batch training helps to neutralize the various influence of the computing errors caused by the random soft errors. In this work, we set batch size to be 16 for the different model retraining in R2F.

\begin{figure*}[tb]
\centering
    \includegraphics[width=0.99\textwidth]{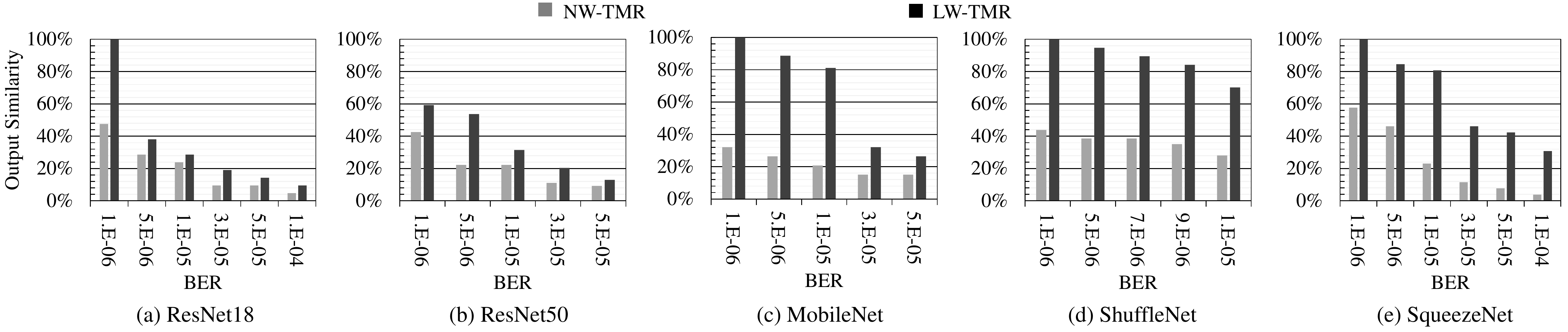}
    \vspace{-0.5em}
    \caption{Last layer output similarity comparison of the different TMR implementations.}
\label{fig:tmrcomp}
\vspace{-1em}
\end{figure*}

TMR is the basis of the R2F optimizations for both the retraining time and the resulting model accuracy. We have two potential TMR implementations, i.e. NW-TMR and LW-TMR evaluated with the output similarity metric. The evaluation result is presented in Figure \ref{fig:tmrcomp}. It shows that LW-TMR shows significantly higher output similarity especially under relatively lower BER. The main reason is that LW-TMR has the computing errors mitigated in layer order of the neural network architecture. The computing errors of the upstream layers are alleviated with TMR before they are passed to the downstream layers. In contrast, NW-TMR has the computing errors passed through the entire neural network and computing errors in upstream neural network layers can aggregate in the downstream neural network layers. Thereby, the computing errors are much larger and the TMR is more likely to fail. When the BER is too high, the majority of the computing errors induced by the soft errors can no longer be mitigated with neither TMR implementations. As a result, the difference narrows down in these cases, which is expected and also roughly exhibits the upper bound of the TMR-based protection.

\begin{figure}
\centering
    \includegraphics[width=0.42\textwidth]{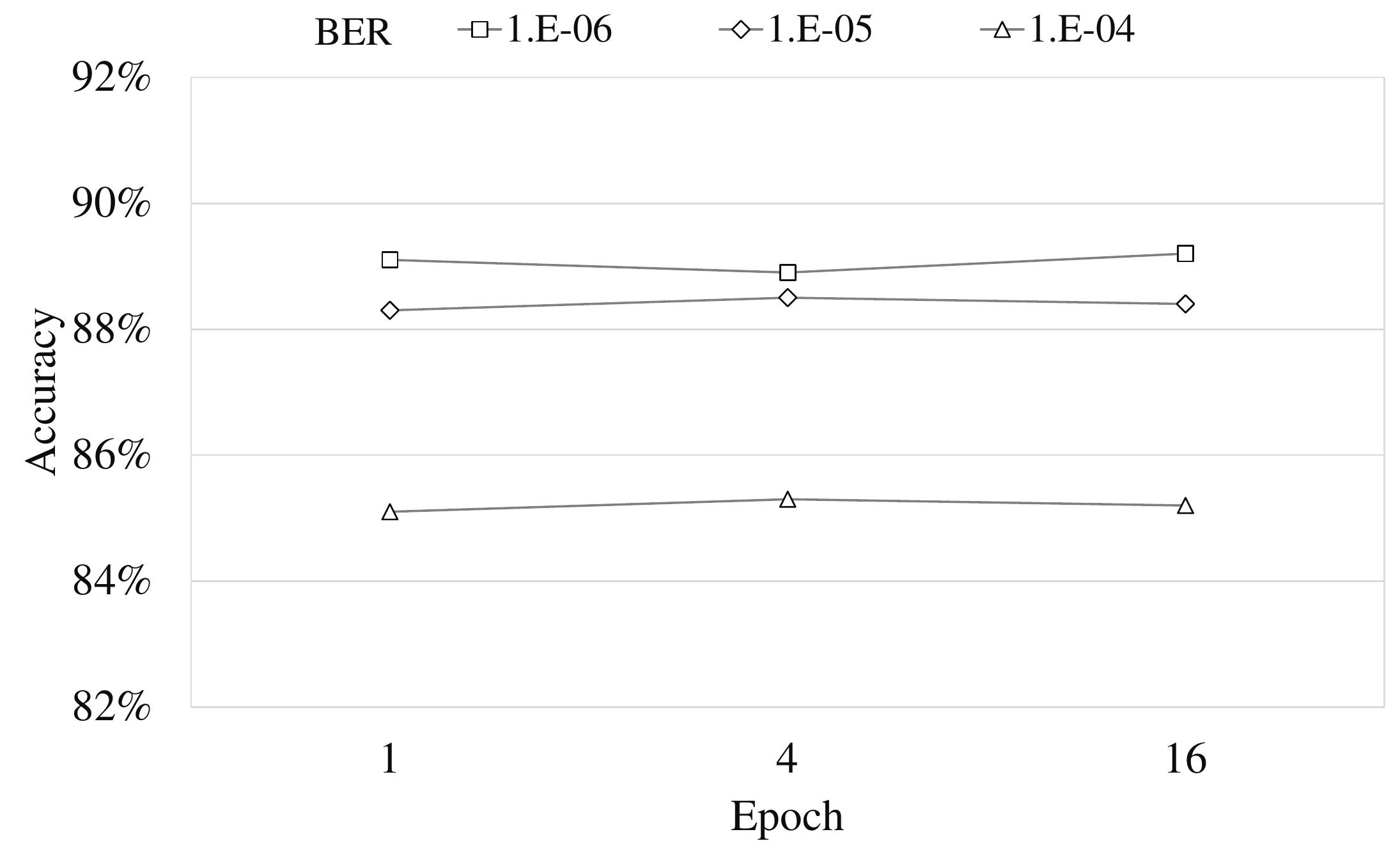}
    \vspace{-0.5em}
    \caption{The influence of different epoch setups on the proposed R2F training.}
\label{fig:epochinfluence-main}
\vspace{-1em}
\end{figure}

Since larger epoch setups typically improve the prediction accuracy of the resulting model. We take ResNet18 as an example and evaluated the model accuracy under different epoch setups. The experiment result is shown in Figure \ref{fig:epochinfluence-main}. It can be observed that the retrained model accuracy shows little improvement given larger epoch. The main reason is that the fault-tolerant retraining is based on a pre-trained model rather than a totally new model. In this case, a single epoch is sufficient according to our experiments. 

\vspace{1em}

\section{Conclusion} \label{sec:conclusion}
Retraining is widely utilized to exploit the inherit redundancy of the neural network models to tolerate the soft errors in AIoT processors, but it is difficult to capture the influence of the soft errors in conventional offline training on GPUs. In this work, we propose R2F, a remote retraining framework, to put the remote AIoT processors in the training loop such that the computing errors caused by the soft errors can be learned with the application data and aware by the resulting models. On top of the basic R2F, we also propose an elastic design trade-off between the model accuracy and the performance penalty with partial TMR optimization to further enhance the retraining. According to our experiments, R2F improves the top-5 model accuracy by 1.93\%-13.73\% with the performance penalty ranging from 0\%-200\%. In addition, we notice that the remote retraining requires a large amount of intermediate data transmission between the AIoT processors and the server, which even dominates the training time due to the limited uplink bandwidth in the AIoT processors. To address the problem, we propose a sparse increment compression approach by taking advantage of the TMR to reduce the data transmission significantly. Our experiment results reveal that the retraining time can be reduced by 38\%-88\% on average depending on the BER.

\section{Acknowledgement}\label{acknowledgement}
This paper is supported in part by the National Key
Research and Development Program of China under grant
2020YFB1600201, and in part by the National Natural Science Foundation of China (NSFC) under grant No.(61902375, 61876173). The corresponding author is Cheng Liu.

\bibliographystyle{IEEEtran}
\bibliography{refs} 



%



\end{document}